\documentclass[twocolumn,preprintnumbers,amsmath,amssymb]{revtex4}
\usepackage{graphicx,color}
\usepackage{amsmath,amssymb,latexsym}

\begin{document}

\title{Noise-induced dynamical phase transitions in
long-range systems}

\author{Pierre-Henri Chavanis}
\affiliation{
Laboratoire de Physique Th\'eorique (IRSAMC), CNRS and UPS,
 Universit\'e Paul Sabatier, F-31062 Toulouse, France}
 
\author{Fulvio Baldovin}
\affiliation{
Dipartimento di Fisica, Sezione INFN, CNISM, and
Universit\`a di Padova,
 Via Marzolo 8, I-35131 Padova, Italy
}

\author{Enzo Orlandini}
\affiliation{
Dipartimento di Fisica, Sezione INFN, CNISM, and
Universit\`a di Padova,
 Via Marzolo 8, I-35131 Padova, Italy
}

\date{\today}

\begin{abstract}
In the thermodynamic limit, the time evolution of isolated long-range
interacting systems is properly described by the Vlasov
equation. This equation admits non-equilibrium dynamically stable
stationary solutions characterized by a zero order parameter. We show
that the presence of external noise sources, like for instance a heat
bath, can induce at a specific time a dynamical phase transition
marked by a non-zero order parameter.
This transition
corresponds to a restoring of the full ergodic properties of the system
and may be used as
a distinctive experimental signature of the existence of
non-equilibrium Vlasov-stable states. 
In particular, we evidence for the first time a regime characterized
by an order parameter pulse.
Our analytical results are
corroborated by numerical simulations of a paradigmatic long-range
model.
\end{abstract}

\maketitle

The study of the nonequilibrium behavior of long-range interacting
systems recently triggered  an intense research activity
\cite{general}.  This is partly a consequence of the
variety of physical situations characterized by long-range
interactions (including gravitational, plasma and nuclear physics,
hydrodynamics, wave-matter interactions,\ldots) \cite{general}, which
makes the topic amenable to interdisciplinary approaches and
cross-fields fertilization. However, in most of the literature in
this area the influence of an external environment that may act as a
thermal bath has not been taken into account. Especially for
laboratory-scale systems, such an influence is expected to be
relevant.  Here we show that the existence of external
noise sources can induce, on long-range systems characterized by
different equilibrium phases, a dynamical phase transition from a
homogeneous Vlasov-stable phase to an inhomogeneous one.  This
dynamical transition corresponds to a restoring of ergodicity realized
by the action of the noise. By extending the linear stability analysis
of the Vlasov equation in order to take into account the influence of
the noise on the system's dynamics, we identify the time $t_c$ at
which the dynamical phase transition occurs.  At $t_c$, in the
thermodynamic limit, phase functions like the order
parameter are not analytic.  We thus provide a ``dynamical
phase diagram'' in which time replaces the role of a
thermodynamic parameter in ordinary equilibrium phase diagrams. 
For specific ranges of initial
energies, the system undergoes a dynamical phase transition even if the
final equilibrium phase is homogeneous.  This behavior corresponds to
an {\it order-parameter pulse} occurring at $t=t_c$. Our analytical
results are corroborated by numerical simulations of a paradigmatic
model, which fully agree with the analytical predictions.

A system is considered ``long-range interacting'' if the interparticle
potential $V$ decays at large distances $r$ slower than $1/r^d$, where
$d$ is the system's spatial dimension. Under these circumstances,
phase transitions have to be addressed within a proper definition of
the thermodynamic limit, which is attained by rescaling the
interaction strength with a function of the system's size in order to
make the energy formally extensive \cite{general} (Kac's prescription)
and by taking the number of particles $N\to\infty$.  In this limit,
the interparticle correlations (due to finite-$N$ effects) are
completely negligible, and the system's time-evolution is correctly
described by the Vlasov equation \cite{general}. We have recently
demonstrated \cite{baldovin_1} that a Hamiltonian thermal bath
\cite{baldovin_2} and a stochastic Langevin bath \cite{chavanis_2}
produce the same equilibrium and nonequilibrium effects when in
contact with a long-range system, a conclusion not obvious {\it a
priori} due to the character of the system's interactions.  Assuming
$d=1$ for simplicity, the time evolution of the one-particle
distribution function $f(x,v,t)$ [normalized such that
$\int_{-\infty}^{+\infty} dx\int_{-\infty}^{+\infty}
dv\;f(x,v,t)=N,\;\forall t\in\mathbb{R}$] is thus conveniently
described by adding to the Vlasov equation a diffusion and a damping
term. 
This
gives the so-called mean-field Kramers' equation \cite{chavanis_2}:
\begin{equation}
\frac{\partial f}{\partial t}
+v\frac{\partial f}{\partial x}
-\frac{\partial \Phi}{\partial x}\frac{\partial f}{\partial v}
=\frac{\partial }{\partial v}
\left(D\frac{\partial f}{\partial v}+\gamma f\,v\right),
\label{eq_kramers}
\end{equation}
where 
$\Phi(x,t)\equiv
\int_{-\infty}^{+\infty} dx^\prime\,
V\left(x-x^\prime\right)
\;\int_{-\infty}^{+\infty} dv\;f(x^\prime,v,t)$, and
$D$ and $\gamma$ are respectively the diffusion and damping
coefficients. The former measures the strength of the noise perturbing
the Vlasov dynamics, whereas the latter quantifies the dissipation to
the external environment. Below, we will assume that these two effects
have the same physical origin, like when the system interacts with an
heat bath. In particular, we suppose that the temperature $T$ of the
bath satisfies the Einstein relation $D=\gamma\,T$ (we work in natural adimensional
units in which the Boltzmann constant and the mass of the particles are equal to one). 
One can
however conceive more general situations in which the noise strength
is not directly linked to the dissipation rate. An interesting recent
example \cite{mukamel_1} is the consideration of a noise term
representing energy-preserving inter-particles fast collision
processes that are superposed to the long-range interactions.  We are
interested in situations in which the potential $V$ is such that there
exist a critical temperature $T_c$ that separates a high-temperature
equilibrium homogeneous phase from a low-temperature inhomogeneous
one. These phases can be described in terms of an order parameter,
$\varphi$, which is zero at and above $T_c$, and non-zero below
$T_c$. Furthermore, we assume that the interparticle potential is even
in the spatial coordinate: $V(-x)=V(x)$.

Let us consider a spatially homogeneous initial condition
$f(x,v,0)=f_0(v)$ that is Vlasov stable. For instance, this could be
the result of a violent relaxation, as in Lynden-Bell's theory
\cite{lb}. Hence, $\partial \Phi/\partial x=0$ ($\varphi=0$)
and the system evolves under the sole effect of the interaction with
the thermal bath according to 
\begin{equation}
\frac{\partial f}{\partial t}
=\frac{\partial }{\partial v}
\left(D\frac{\partial f}{\partial v}+\gamma f\,v\right).
\label{eq_df_1}
\end{equation}
Notice that for $\gamma=D=0$ (microcanonical setup), we recover the
Vlasov stationarity condition $\partial f/\partial t=0$, which would
last for infinite time in this thermodynamic limit. In
contrast, for $\gamma\neq0$, the distribution will change with time
due to the interaction with the bath.  Equation (\ref{eq_df_1})
remains valid only as long as $f(v,t)$ is dynamically stable. If, at
time $t=t_c$, it becomes Vlasov unstable, a dynamical phase transition
occurs from the homogeneous $\varphi=0$ phase to an inhomogeneous
$\varphi\neq0$ phase \cite{finite_N}.
The time $t_c$ can be identified by
performing a linear (Landau) stability analysis \cite{general,cd} of
the Vlasov equation for the $f(v,t)$ which is solution of
Eq. (\ref{eq_df_1}) at time $t$ (in this analysis, $f(v,t)$ is
regarded as a function of $v$, the time $t$ being fixed
adiabatically).  While a general $f(v,t)$ can be studied within the
Landau formalism, in the following we will restrict ourselves, for the
sake of simplicity, to distributions with a single maximum at $v=0$.
In this case, one then shows \cite{general,cd} 
that $t_c$ is the minimum time $t$ for which the equation 
\begin{equation}
1-{\widetilde V(k)}\int_{-\infty}^{+\infty}\frac{\frac{\partial f}{\partial v}(v,t)}{v} \, dv=0
\label{eq_tc}
\end{equation}
admits a solution for at least a $k\in\mathbb{R}$, $k\neq0$. This
expresses marginal stability.  Here, $\widetilde V(k)\in\mathbb{R}$ is the Fourier transform of the
potential: $\widetilde V(k)\equiv\int_{-\infty}^{+\infty}
dx\;e^{ikx}\;V(x)$.  In order to find $t_c$, we can analytically
solve Eq. (\ref{eq_df_1}) for a given uniform initial condition
$f(x,v,0)=f_0(v)$, plug the result in Eq. (\ref{eq_tc}), and
invert the solution corresponding to the minimum time.

The Green function of Eq. (\ref{eq_df_1}) is 
\begin{equation}
W(v,t|v_0,t=0)=\frac{\exp\left[-(v-a(t)\;v_0)^2/2\;\sigma^2(t)\right]}
{\sqrt{2\;\pi\;\sigma^2(t)}},
\end{equation}
where $a(t)\equiv e^{-\gamma t}$, and 
$\sigma^2(t)\equiv T\left(1-e^{-2\gamma t}\right)$.
Formally, the expression for $f(v,t)$ is thus
\begin{equation}
f(v,t)=\int_{-\infty}^{+\infty} dv_0\;W(v,t|v_0,t=0)\;f_0(v_0).
\label{eq_formal}
\end{equation}
Since the time evolution can be expressed in
adimensional units through the mapping $t\mapsto\gamma t$, 
below we implicitly adopt this
convention. 
Our problem can be solved in Fourier space. 
First we take the
Fourier transform in the velocities, 
\mbox{$\widetilde{[\cdot]}(w)\equiv\int_{-\infty}^{+\infty}
dv\;e^{iwv}\;[\cdot](v)$}, of Eq.(\ref{eq_formal}), to get
\begin{equation}
\widetilde f(w,t)=\widetilde f_0\left(a(t)\;w\right)\;
e^{-\sigma^2(t)\;w^2/2}.
\end{equation}
Recalling the expressions for $a(t)$ and $\sigma^2(t)$, it is clear
that the short-time behavior
of $\widetilde f(w,t)$ is dominated by the initial condition
$\widetilde f_0\left(w\right)$ (e.g., by the initial system's energy), 
while its long-time behavior is determined by the
heat bath temperature.
Inserting the Fourier representation of
$f(v,t)$ in Eq. (\ref{eq_tc}), we get that 
the transition time is equivalently identified
by the zeroes of 
the function  
\begin{equation}
I(t,k)\equiv1+\frac{\widetilde V(k)}{2}\int_{-\infty}^{+\infty} dw\;|w|\;
\widetilde f_0\left(a(t)\;w\right)\;
e^{-\sigma^2(t)\;w^2/2},
\end{equation}
for some $k\neq0$.

The analysis outlined so far is valid under the following general
assumptions: (i) One-dimensional system;
(ii) Kac's thermodynamic limit; (iii) Symmetric potential; 
(iv) Spatially homogeneous distribution function $f(v,t)$ with a
single maximum at $v=0$.  
We now specify the above general formulas to the Hamiltonian Mean Field
(HMF) model, a long-range interacting system that has been widely used
in the recent literature as a paradigmatic case of study \cite{general}.
The HMF model,  
which can be
thought of as a set of $N$ globally coupled rotators or
$XY$-spins, is described by the Hamiltonian
\begin{eqnarray}
H= \sum_{i=1}^N\frac{v_i^2}{2}+
\frac{1}{N}\sum_{i<j}\left[1-\cos\left(\theta_i-\theta_j\right)\right],
\label{eq_hmf}
\end{eqnarray} 
where $\theta_i\in[0,2\pi[$ is an angular coordinate, $V(\theta)\equiv
(1-\cos\theta)/N$ is the (cosine) potential, and the order parameter
$\varphi$ amounts to the system's magnetization:
\mbox{$\varphi\equiv\left|\sum_{i=1}^N(\cos x_i,\sin x_i)\right|/N$}.
Notice that, following the Kac's prescription, the interaction strength has been rescaled by a factor $1/N$.  
In this way the total energy of the system, 
$E=\sum_{i=1}^N v_i^2/2+N(1-\varphi^2)/2$,
becomes extensive, and we can consistently introduce the specific
energy $e\equiv E/N$.  
Once coupled with a Langevin thermal bath as in Eq. (\ref{eq_kramers}),
the system has also been named Brownian Mean Field (BMF) model
\cite{chavanis_2}. 

\begin{figure} 
\includegraphics[width=0.65\columnwidth]{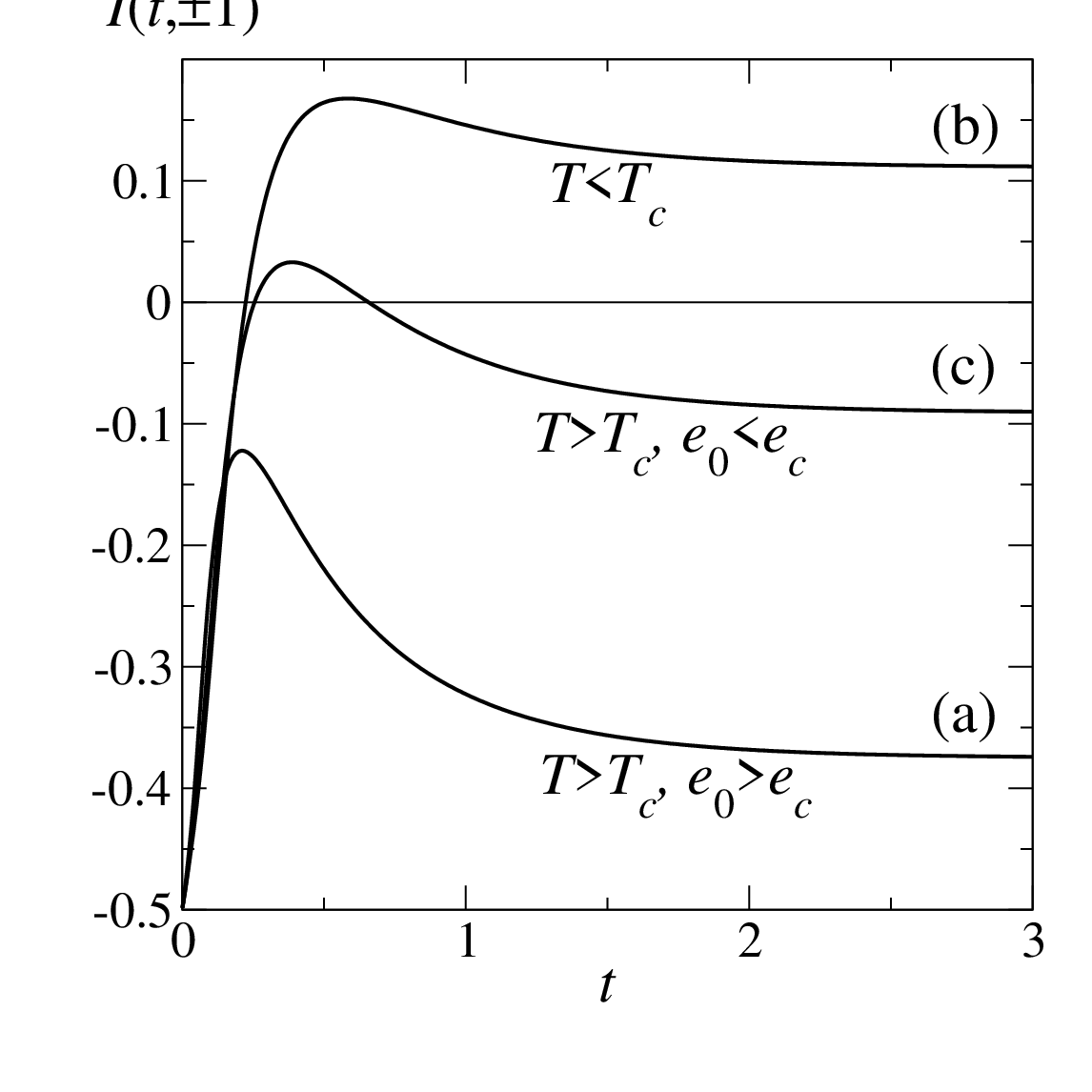}
\caption{
  Stability analysis in the presence of external noise. Depending on
  the initial energy and on the temperature, an initially stable
  uniform state may stay uniform forever (a); becomes non-uniform at
  $t=t_c$ and remains non-uniform (b); becomes non-uniform at $t=t_c$
  and then becomes uniform again (c).  }
\label{fig_integral}
\end{figure}

\begin{figure} 
\includegraphics[width=0.65\columnwidth]{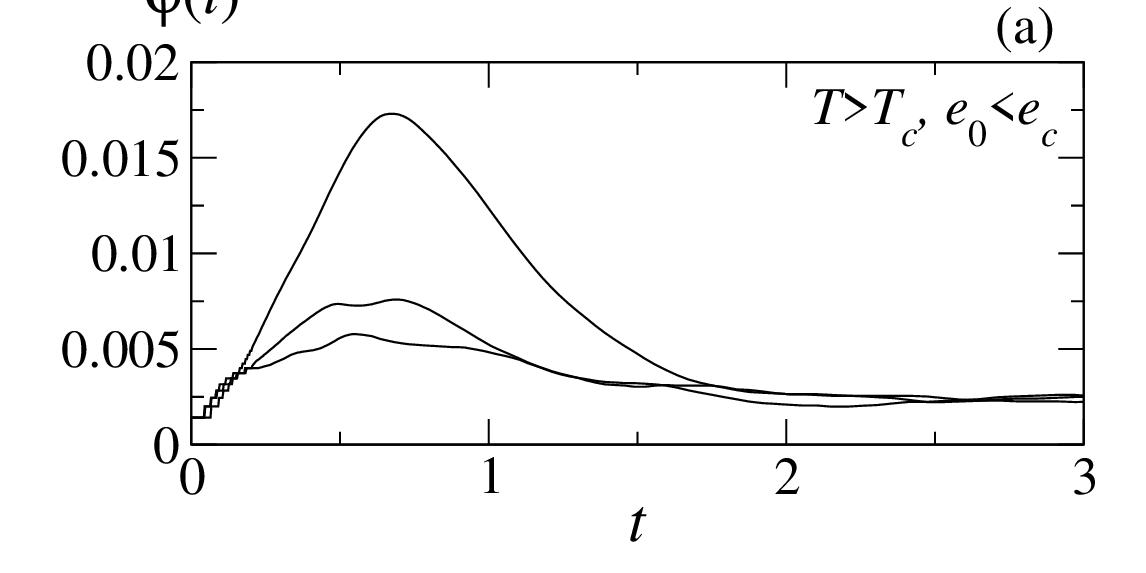}\\
\vspace{0.8cm}
\includegraphics[width=0.65\columnwidth]{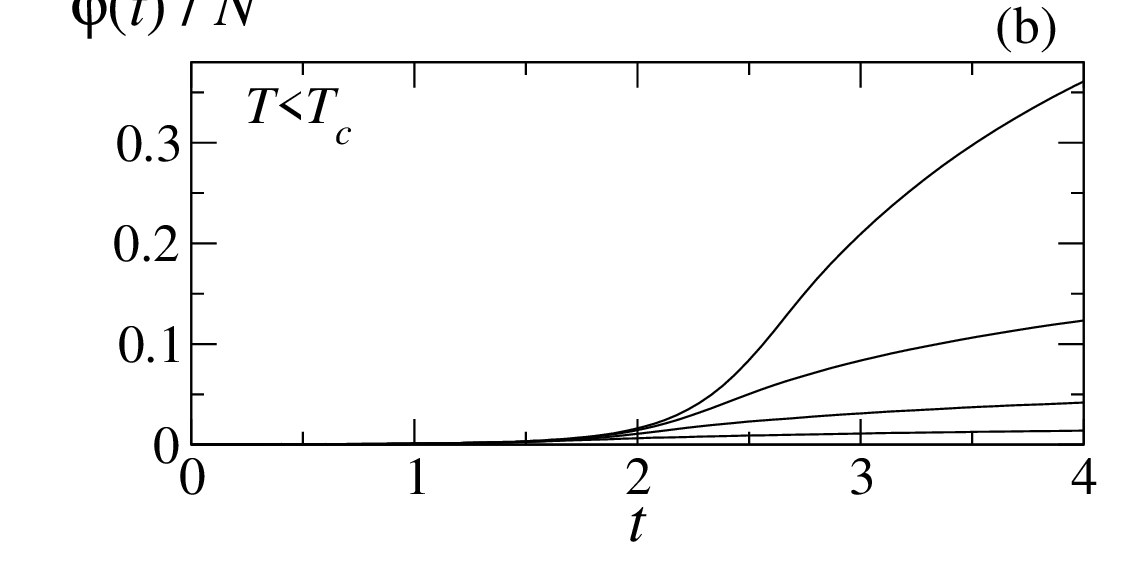}
\caption{
  (a) Magnetization pulse revealing a dynamical phase
  transition. Different curves are obtained for 
  $T=0.6>T_c$, $N=10^6$, averaged over $20$ runs with initial 
  conditions as in Eq. (\ref{eq_f0_1}).  
  From top to bottom $e_0=0.60,\;0.63,\;0.65<e_c(0.6)=0.662\ldots$. 
  (b) Finite-size analysis for $T=0.40<T_c$ and $e_0=1.0$. From top to bottom the lines refer to 
  $N=10^7,\;10^6,\;10^5,\;10^4$.
}
\label{fig_pulse}
\end{figure}

To begin with, we consider uniform initial
conditions: 
\begin{equation}
f_0(v)=
\frac{N}{4\pi\overline v}\;[\vartheta(v+\overline v)-\vartheta(v-\overline v)]
\label{eq_f0_1}
\end{equation}
($\vartheta$ being the Heaviside step function), where
$\overline v=\sqrt{6(e_0-1/2)}$ and $e_0$ is the initial specific energy
of the system. 
In Fourier space we thus have
$\widetilde f_0(w)=N\sin(\overline vw)/2\pi\overline vw$, 
$\widetilde V(k)=\pi\left(2\delta_{k,0}-\delta_{k,1}-\delta_{k,-1}\right)/N$, and we are led
to examine the zeroes of \cite{note_erfi}
\begin{equation}
I(t,\pm1)=1-\frac{1}{2a(t)\overline v}\;
\int_0^{+\infty} dw\sin\left(a(t)\overline v w\right)
e^{-\sigma^2(t)\;w^2/2}.
\end{equation}
If
$e_0<e^*\equiv 7/12$, $I(0,\pm1)>0$, which means that at
$t=0$ $f_0(v)$ is dynamically unstable \cite{general,cd}.  Since our
analysis applies to states which are initially stable, in the
following, we will restrict ourselves to cases with
$e_0>e^*$. According to the values $e_0$ and $T$, there are three
possible qualitative behaviors of $I(t,\pm1)$, as depicted in
Fig. \ref{fig_integral}. In Fig. \ref{fig_integral}a, $I(t,\pm1)<0$
$\forall t>0$, and the system magnetization $\varphi$ is always zero
(no dynamical phase transition occurs). Correspondingly,
$t_c=+\infty$. This happens if $T$ is above the critical value \cite{general}
$T_c=1/2$ [$I(+\infty,\pm 1)<0$] and if the initial
energy $e_0$ is sufficiently high. An alternative is that $I(t,\pm1)$
crosses zero for a single value $t=t_c$
(Fig. \ref{fig_integral}b). For $t<t_c$ $\varphi$ is zero, whereas for
$t>t_c$ the system becomes and remains magnetized. This occurs for
$T<T_c$. There is however a third possibility: $I(t,\pm1)$ may cross
zero twice (Fig. \ref{fig_integral}c). This occurs for $T>T_c$ and if
$e_0$ is below a critical value $e_c(T)$, which in general also
depends on the shape of $f_0(v)$. The first zero corresponds to a
dynamical phase transition occurring at $t=t_c$, which drives the
system to $\varphi>0$. Since $T>T_c$, the system must eventually come back to zero
magnetization in order to reach equilibrium. Under these conditions,
one observes a ``magnetization pulse'' in the behavior of $\varphi(t)$
(Fig \ref{fig_pulse}a).  Notice that we cannot interpret the second
zero of $I(t,\pm1)$ as the second transition time, since our stability
analysis is only valid under the assumption of spatially homogeneous
conditions, which breaks at $t=t_c$. However, the
system eventually goes back to $\varphi=0$ to equilibrate with the
thermal bath. Notice also that in this thermodynamic limit
description, whenever the dynamic phase transition occurs,
$\varphi(t)$ is non-analytic at $t=t_c$. This can be
checked numerically with the BMF model, recalling that 
$\varphi$ in a homogeneous state scales with the system size $N$ as
$\varphi(N)\sim1/\sqrt{N}$ (in the inhomogeneous phase, $\varphi\sim 1$). 
Rescaling by $\sqrt{N}$ the curves
obtained at different $N$'s,  we indeed find
that all the curves collapse for $t<t_c$ and develop the same
non-regular behavior at $t=t_c$ (Fig. \ref{fig_pulse}b).

\begin{figure} 
\includegraphics[width=0.65\columnwidth]{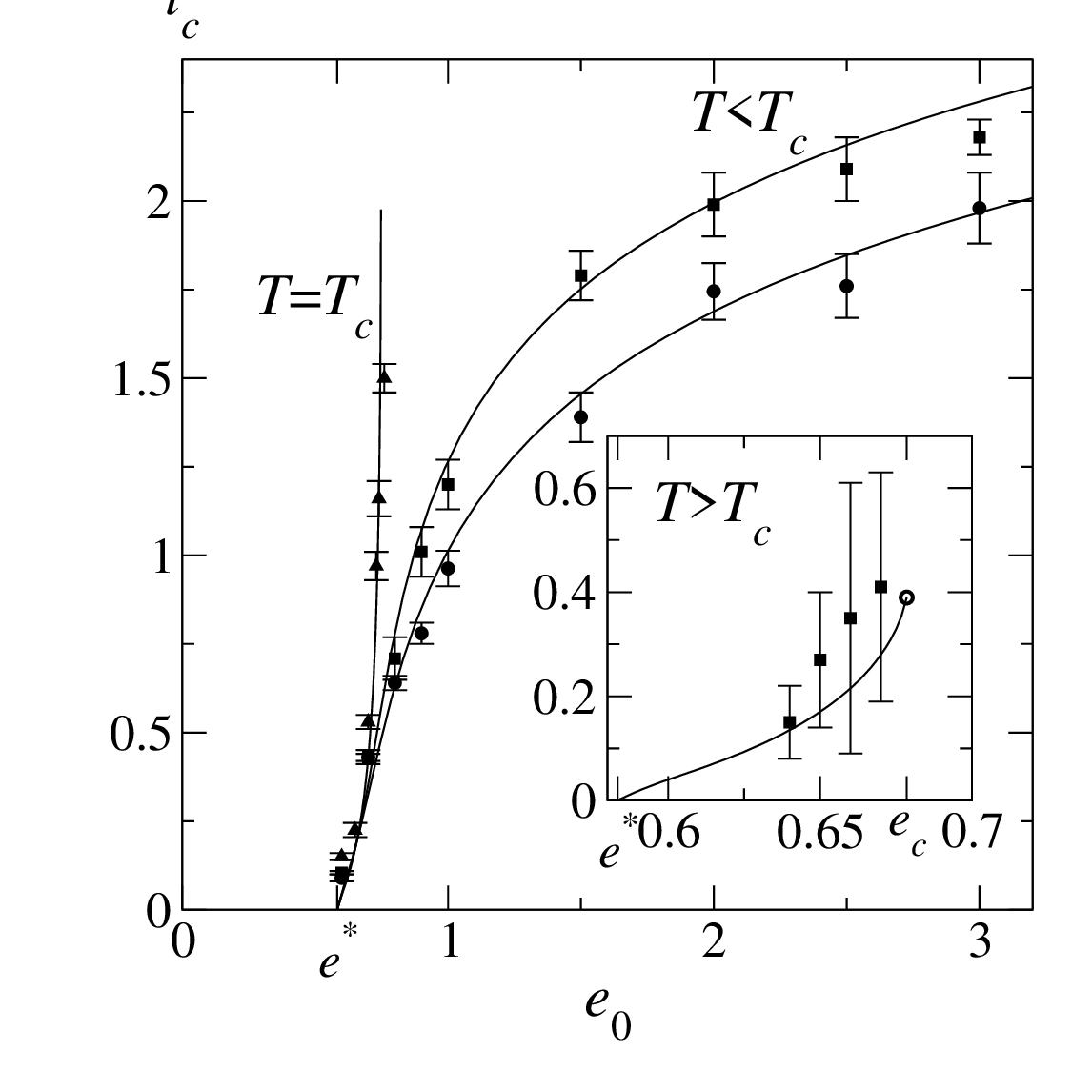}
\caption{
  Phase diagram of the dynamical phase transitions. At $T<T_c=1/2$, $t_c$ is
  always finite (plots refer to $T=T_c=0.5$, $T=0.45$, $T=0.4$). 
  Inset: At $T>T_c$ the curves end at a critical value
  $e_c(T)$ [plot refers to $T=0.55$, $e_c(0.55)=0.584\ldots$]. 
  Above this energy, $t_c$ becomes infinite. Points and
  error bars refer to simulations with $N=10^7$, averaged over $20$
  runs.  
}
\label{fig_phase_dia}
\end{figure}

All the previous findings can be summarized in a dynamical phase
transition diagram representing $t_c$ vs $e_0$, for different values
of $T$ (solid lines in Fig. \ref{fig_phase_dia}). 
If $T<T_c$, a dynamical
phase transition always occurs and, correspondingly,
$0<t_c<+\infty\;\;\forall e_0>e^*$.  For $T\geq T_c$, the transition
line ends at a critical value $e_c(T)$: if $e_0>e_c(T)$, $t_c$ becomes
infinite and the dynamical phase transition does not occur anymore. In
the range $e^*<e_0\leq e_c(T)$, the dynamical evolution of $\varphi$
corresponds in fact to a ``pulse'' starting at $t=t_c$ (see
Fig. \ref{fig_pulse}a).  Fig. \ref{fig_phase_dia} also reports the
results of numerical simulations of the BMF model \cite{chavanis_2},
which fully agree with the analytical curves.
It is also possible to obtain 
two asymptotic expressions for $t_c$ in a closed form
\cite{preparation}: 
for $T<T_c$ and $e_0\gg1$, $t_c$ grows logarithmically as 
\begin{equation}
t_c\sim \frac{1}{2}\ln\left[\frac{12e_0-6(1+T)}{6T(1-T/T_c)}\right].
\end{equation}
For $T\gg T_c$ and $e_0\to e^{*\,+}$, $t_c$ goes to zero linearly as
\begin{equation}
t_c\sim \frac{6}{\gamma}\frac{e_0-e^*}{1+T/\left(6e_0-3\right)}.
\end{equation}

We have also checked the validity of our conclusions for a class of
(spatially uniform) initial distributions $f_0(v)$ wider than the
water-bag, namely: 
\begin{eqnarray}
f_0(v)&=&
\frac{N}{2\pi}\;
\left[
\kappa+\left(\frac{\pi}{4\overline v}-\frac{\pi \kappa}{2}\right)
\cos\left(\frac{\pi v}{2\overline v}\right)
\right]\nonumber\\
&&\times[\vartheta(v+\overline v)-\vartheta(v-\overline v)].
\label{eq_f0_2}
\end{eqnarray}
Here, $0\leq \kappa\leq 1/2\overline v$ is a form factor with dimension of
inverse velocity, and $\overline v$ depends now on both $e_0$
and $\kappa$. For $\kappa=1/2\overline v$, Eq. (\ref{eq_f0_2}) reduces to
Eq. (\ref{eq_f0_1}), whereas for $\kappa=0$ $f_0(v)$ becomes a bump-like
distribution that touches the $v$-axis. 
Eq. (\ref{eq_f0_2}) can be understood as a way of modeling the action
of a velocity-selector with milder cutoff effects than the water-bag,
which are parametrized by the form-factor $\kappa$.
With respect to this class of initial conditions, we have found that
$t_c(\kappa)$ is in general a monotonically decreasing function, meaning
that for a given $e_0$ the water-bag initial conditions correspond to
the most stable distribution of the class. 
However, all the qualitative features of Fig. \ref{fig_phase_dia}
are confirmed for any value of allowed $\kappa$.

The dynamical evolution of long-range interacting systems is dominated
by a global interaction which erases the interparticles correlations when 
$N\rightarrow +\infty$. 
As a consequence, the system gets easily trapped in non-equilibrium
states which correspond to the occupation of a small fraction of the
available system's phase space \cite{baldovin_1}.
We have shown that the action of a reservoir or of an external
environment, renders these non-equilibrium states dynamically
unstable, restoring the full (ergodic) occupation of the system's
phase space. 
Correspondingly, the system manifests a dynamical phase transition,
which can be detected by monitoring the time evolution of the order 
parameter. 
The transition time $t_c$ can be exactly identified by extending the
Landau stability analysis to take into account the influence
of the heat bath. The result is a dynamical phase diagram, in which
time replaces the role ordinarily played by a thermodynamic parameter.
The dynamical transition may occur even if the external temperature is
above $T_c$, and under these circumstances it is revealed by a peak in
the time evolution of the order parameter. 
We have studied situations in which the external noise satisfies a
fluctuation-dissipation relation which enables the definition of the
heat-bath temperature. However, it is also possible to extend the Landau
stability analysis to more general cases.  
This could be important to reproduce experimental conditions in which
many sources of noise that do not necessarily satisfy a
fluctuation-dissipation relation act on the system.  
The generality of our approach and the robustness of our 
results under different initial conditions suggest that
these dynamical phase transitions should be common and marked enough,
to overcome experimental test. 
In particular, a pulse in the time evolution of the
order parameter may be used as a signature of the existence of
Vlasov stable out-of-equilibrium states in experiments with
long-range interacting systems.

{\bf Acknowledgments} 
FB and EO acknowledge fruitful discussions with A.L. Stella.


\begin{thebibliography}{99}
\bibitem{finite_N}
A similar loss of stability has been reported in \cite{campa}. In
that case, the system is isolated and the distribution function
changes due to finite $N$ effects (``collisions'') until it becomes
Vlasov unstable. Because of the finiteness of $N$, such a
stability-loss relates a Vlasov-stable regime in which
$\varphi\sim1/\sqrt{N}$ to the equilibrium phase (see also
\cite{mukamel_1}).
\bibitem{note_erfi}
The integral can be expressed in terms of ${\rm Erfi}(x)$.
\end{thebibliography}
\end{document}